\documentclass[review,12pt,authoryear]{elsarticle}
\usepackage{cmap}
\usepackage[utf8]{inputenc}
\usepackage[T1]{fontenc}

\usepackage{url}
\usepackage{amsmath}
\usepackage{graphicx}
\usepackage{geometry}
\journal{Tectonophysics}

\begin{document}

\begin{frontmatter}

\title{Cross-Sectional and Spatio-Temporal Analysis of Seismicity Parameters in the Zagros Orogenic Belt: Insights into Crustal Stress Distribution and Seismic Hazard}

\author[1]{Muhammed Hossein Mousavi}
\author[2]{Marjan Tourani}
\author[3]{Amir Talebi}
\author[4]{Hamzeh Mohammadigheymasi}
\author[5]{Ivan Koulakov}

\address[1]{Department of Physics Education, Farhangian University, P.O. Box 14665-889, Tehran, Iran}
\address[2]{Department of Geological Engineering, Ankara University, Tectonics Research Group, Ankara, T\"{u}rkiye}
\address[3]{Department of Physics, Faculty of Science, Arak University, Arak 384817758, Iran}
\address[4]{Atmosphere and Ocean Research Institute (AORI), the University of Tokyo, Kashiwa, Japan}
\address[5]{Skolkovo Institute of Science and Technology (Skoltech), Moscow, Russia}

\begin{abstract}
The Zagros Orogenic Belt, formed by the ongoing collision between the Arabian and Eurasian plates, is one of the most seismically active continental collision zones in the world, hosting a large proportion of Iran's recorded seismicity and posing significant earthquake hazard to densely populated regions. Despite its tectonic importance, a comprehensive and spatially integrated analysis of key seismotectonic parameters across the entire belt using recent instrumental data remains limited. In this study, the IRSC earthquake catalog (2006--2024) was used to construct a homogeneous seismic dataset for the Zagros Orogenic Belt, and the spatial distribution of seismicity was analyzed together with key seismotectonic parameters, including $b$-value, fractal dimension ($D_c$-value), and differential stress ($\sigma_1 - \sigma_3$), to evaluate stress variations, clustering behavior, and earthquake hazard potential. The obtained $b$-value of $0.81 \pm 0.01$ indicates a dominantly high-stress crustal regime, while low $b$-value zones ($0.4$--$0.7$) are systematically concentrated along major fault systems such as the Mountain Front Fault, High Zagros Fault, and Main Zagros Reverse Fault, reflecting strong tectonic loading and segmented deformation. Cross-sectional analyses reveal that low $b$-values ($<0.6$) are mainly confined to the upper $\sim$10~km of the crust, indicating that seismic energy release is concentrated within the shallow brittle layer. $D_c$-values ranging from 1.0 to 2.05, with dominant high values ($\sim$1.5 and above), show a strong spatial correspondence with low $b$-value zones, indicating highly heterogeneous and structurally complex fault systems. This heterogeneity persists with depth, where high $D_c$-values dominate the upper crust, while localized decreases suggest more coherent rupture behavior in deeper segments. The differential stress distribution (100--520~MPa, predominantly $\geq$520~MPa) confirms a region-wide high tectonic loading associated with the Arabia--Eurasia collision, and exhibits a clear inverse relationship with $b$-value, where high-stress zones correspond to low $b$-value and high $D_c$-value areas. Cross-sectional results consistently support this coupling between all parameters, indicating that they delineate the same critically stressed fault-controlled regions. Overall, the integrated analysis demonstrates that seismic hazard in the Zagros Orogenic Belt is controlled by shallow, highly stressed, and structurally complex fault segments, which represent the most likely sources of future moderate to large earthquakes. These findings emphasize the critical need for continuous seismic monitoring, detailed hazard assessment, and the implementation of effective mitigation and preparedness strategies across the region.
\end{abstract}


\end{frontmatter}

\section{Introduction}\label{sec:intro}
The Zagros Orogenic Belt represents one of the most seismically active tectonic regions in the world and developed as a consequence of the continental collision between the Arabian Plate and the Turkish and Iranian continental blocks following the closure of the Neo-Tethys Ocean from the Late Cretaceous to the Miocene \citep{berberian1981,sengor1984,sharland2001,jassim2006,doski2022}. Consequently, the Zagros Mountains constitute the most seismically active tectonic belt in Iran, hosting more than half of the country's recorded earthquakes, while their structural complexity, characterized by major strike-slip faults and extensive fold-and-thrust systems, makes the Zagros one of the most seismically active regions in the world \citep{mirzaei1998,talebi2020}. Earthquakes in this region occur abruptly due to sudden fault slip, and seismic activity has been particularly intense in recent years. Between 2006 and 2022, the study area experienced 12 earthquakes with magnitudes greater than 6. The most powerful event, the $M_w$ 7.3 earthquake of 12 November 2017, ranks among the strongest recorded in Iran. The $\sim$30-second rupture caused widespread damage across western Iran and was felt as far as central regions, including high-rise buildings in Tehran. The most severe destruction occurred in Sarpol-e Zahab, Qasr-e Shirin, Javanrud, Salas-e Babajani, and Gilan-e Gharb, resulting in 430 fatalities and 7,460 injuries according to IRNA. Consequently, numerous studies have focused on seismicity characteristics, probabilistic seismic hazard assessment (PSHA), and integrated fault- and catalog-based hazard models across Iran, particularly within the Zagros \citep{hatzfeld2003,madahizadeh2016,farahi2017,nazarinezhad2024,mousaviyan2022,dolatabadi2021combined}.

A key approach for evaluating seismic conditions, hazards, and regional tectonics involves analyzing seismotectonic parameters such as the $b$-value of the Gutenberg--Richter relationship, the fractal dimension ($D_c$-value), and differential stress ($\sigma_1-\sigma_3$) \citep{gutenberg1944,oncel2002,polat2008,ozturk2015,scholz2015,hussain2020,mousaviyan2022}. The $b$-value reflects the slope of the magnitude--frequency distribution \citep{gutenberg1944}. Earlier studies have reported that $b$-values generally vary between 0.4 and 2.0 \citep{wiemer2002}, whereas both worldwide and regional-scale investigations consistently suggest an average $b$-value of approximately 1.0 \citep{frohlich1993,kagan1999,mishra2007,wech2010,elisa2014}. Lower $b$-values are commonly associated with the potential for larger earthquakes \citep{hussain2020,chiba2022,tourani2024,mousavi2026c}. Numerous studies have documented $b$-value variations preceding major earthquakes \citep{hanks1979,smith1981,wyss1997,cao2002,kijko2002,nuannin2012,elisa2014,ahmed2016,kulhanek2018,xie2019,gui2019,godano2022,lacidogna2023,hisyam2024}. Additionally, a relationship can be observed between $b$-values and the structural fault blocks within the study area. Several studies have demonstrated that $b$-values are controlled by earthquake focal mechanisms and faulting styles \citep{schorlemmer2005,gulia2010,scholz2015,hussain2020,lewerissa2021}. Collectively, these studies reveal that different fault types exhibit distinct $b$-value ranges, where low to medium $b$-values (0.6--0.9) are generally associated with thrust and strike-slip fault blocks, whereas higher values ($>$1.0) are typically correlated with normal fault systems.

Moreover, both earthquake statistical analyses and laboratory experiments on small-scale rock specimens have consistently demonstrated a well-established inverse relationship between $b$-value and differential stress \citep{amitrano2003,scholz2015}. It is widely recognized that $b$-values exhibit a negative correlation with crustal differential stress ($\sigma_1-\sigma_3$), meaning that $b$-values tend to decrease as differential stress increases \citep{elisa2014,scholz2015,spada2013}. Low $b$-values observed in the vicinity of active faults are commonly interpreted as indicators of possible asperities or locked fault patches \citep{tormann2014,mousavi2026b}. By jointly analyzing $b$-values and differential stress, we can infer the stress state and mechanical condition of the crust, where low $b$-values indicate zones of high differential stress and potential strain accumulation, while higher $b$-values suggest lower stress levels and more heterogeneous deformation, thereby allowing the identification of critically stressed fault segments and potential earthquake-prone areas.

In addition, the fractal dimension ($D_c$-value) is a powerful parameter for characterizing the spatial complexity of seismicity because it quantifies the degree of clustering and heterogeneity in earthquake distributions, which are controlled by stress and fault geometry \citep{grassberger1983}. $D_c$-values provide important insights into the structural organization and deformation patterns of seismically active regions. Lower $D_c$-values indicate stronger clustering, whereas values near 1 reflect linear alignment along active faults, and values near 2 represent uniform epicenter distribution in two-dimensional space \citep{grassberger1983,roy2011,tatar2011}. Previous studies have applied $b$-value, differential stress, and fractal analysis across various regions of Iran \citep{firoozfar2019,mousaviyan2022,sabahi2024,tourani2024,mousavi2025b,mousavi2026b,mousavi2026integrating}.

The aim of this study is to investigate the seismotectonic characteristics, spatial clustering behavior, stress distribution, and earthquake hazard potential of the Zagros Orogenic Belt through the integrated analysis of $b$-value, differential stress ($\sigma_1-\sigma_3$), and fractal dimension ($D_c$-value). By combining these complementary seismic parameters, this study seeks to identify critically stressed fault segments, characterize the spatial complexity of seismicity, and improve the understanding of the relationship between fault geometry, stress accumulation, and earthquake occurrence within one of the most tectonically active regions of the Middle East. In addition to spatial analyses, cross-sectional investigations of these parameters were also performed to evaluate their depth-dependent variations and to better constrain the subsurface geometry of active fault systems and seismogenic zones within the region.
\section{Geology and Tectonic Setting}\label{sec:tectonics}
\begin{figure*}[ht!]
    \centering
    \includegraphics[width=\textwidth]{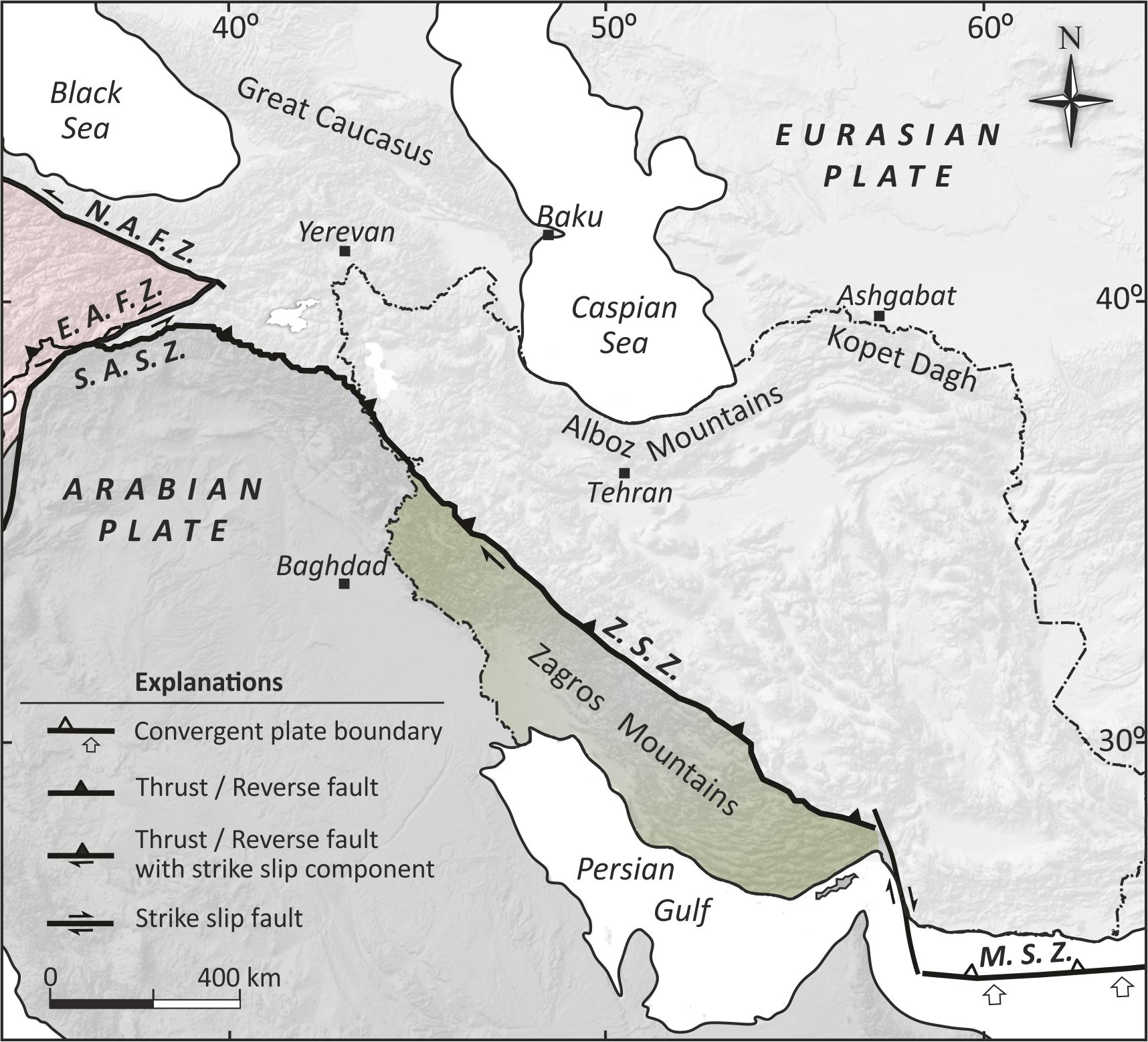}
    \caption{Tectonic setting and geographic position of the Zagros Belt within the regional plate boundary framework. Abbreviations: East Anatolian Fault Zone (E.A.F.Z.), North Anatolian Fault Zone (N.A.F.Z.), South Anatolian Suture Zone (S.A.S.Z.), Zagros Suture Zone (Z.S.Z.), Makran Subduction Zone (M.S.Z.)}
    \label{fig:tectonic_setting}
\end{figure*}

The Zagros province lies inside the Zagros Orogenic Belt, among the Iranian Plateau's megascale tectonic provinces. The Zagros Mountains extend for nearly 2,000~km from eastern T\"{u}rkiye in the northwest to the Makran subduction zone in southeastern Iran and developed as a result of the closure of the Neo-Tethys Ocean followed by the ongoing collision and convergence between the Arabian and Eurasian plates, a process that has produced intense tectonic shortening and the formation of the Zagros fold-thrust belt since the late Cenozoic (Fig.~\ref{fig:tectonic_setting}) \citep{berberian1981,sengor1984,mouthereau2012,talebi2020}
. The belt is characterized by compressional tectonics related to the ongoing convergence between the Arabian and Eurasian plates (Fig.~\ref{fig:tectonic_setting}). According to \citet{demets2010}
, the Arabian Plate moves northward relative to the Eurasian Plate at a rate of approximately 31~mm/yr, although several GPS-based studies have proposed slightly lower convergence rates of around 21~mm/yr \citep{mcclusky2000,vernant2004}
.

Geologically, the Zagros has undergone complex deformation processes that led to the development of numerous parallel tectonic features. The Zagros collision zone can be divided into three main tectonic zones from the southwest to the northeast, crosswise to the principal trend: (1)~the Zagros Fold-Thrust Belt (Z.F.T.B.), (2)~the Zagros Imbricate Zone (Z.I.Z.), and (3)~the Urumieh-Dokhtar Magmatic Arc (U.D.M.A.). Moreover, the Z.F.T.B. can be divided into two structural regions: the High Zagros Belt (H.Z.B.) and the Simple Fold Belt (S.F.B.) \citep{bahroudi2003,mcquarrie2004}
. The Zagros has also been subdivided into five parallel structural domains from southwest to northeast, including the Mesopotamian-Persian Gulf foreland basin, the Simply Folded Belt, the Imbricated Zone, the metamorphic and magmatic Sanandaj-Sirjan Zone, and the Urumiyeh-Dokhtar Magmatic Arc \citep{verges2011,saber2023}
.

\begin{figure*}[ht!]
    \centering
    \includegraphics[width=\textwidth]{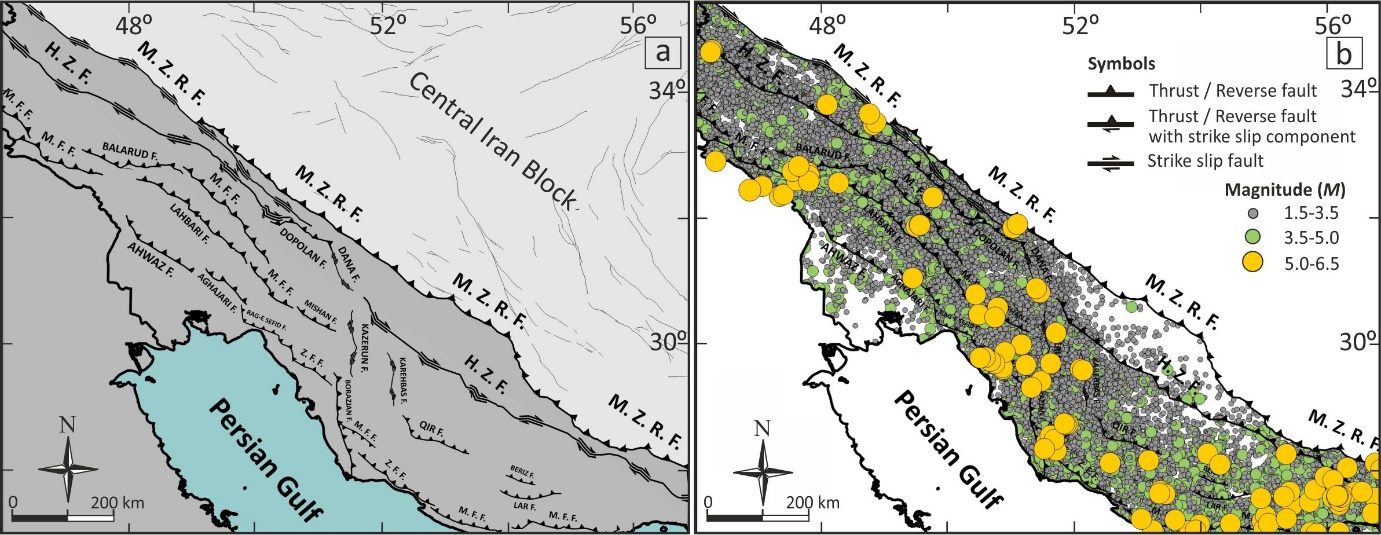}
    \caption{(a) Tectonic framework of the Zagros Mountains illustrating the principal structural domains and major active faults. Abbreviations: Main Zagros Reverse Fault (M.Z.R.F.), High Zagros Fault (H.Z.F.), Zagros Foredeep Fault (Z.F.F.), and Mountain Front Fault (M.F.F.). (b) Seismicity map of the IRSC catalogue for 2006--2024.}
    \label{fig:seismicity_map_zagros}
\end{figure*}

The Zagros Mountains are divided into several NW--SE trending structural zones bounded by major active faults, including the High Zagros Fault (H.Z.F.), the Mountain Front Fault (M.F.F.), and the Main Zagros Thrust (M.Z.T.), which is considered to represent the suture zone of the Arabian and Iranian tectonic plates (Fig.~\ref{fig:seismicity_map_zagros}a). These fault-controlled zones are aligned with the regional collision direction and reflect the ongoing continental convergence between the Arabian and Eurasian plates \citep{berberian1995,agard2011}
. This NW--SE trending fold-and-thrust belt represents one of the youngest and still actively deforming continental collision zones on Earth. It extends for approximately 1500~km, from the Southeast Anatolian Thrust Zone in southeastern T\"{u}rkiye to the Minab Fault region north of the Strait of Hormuz in southern Iran \citep{berberian1981}
. Among these structures, the Mountain Front Fault (M.F.F.) is particularly prominent, characterized by clear morphotectonic expressions, significant topographic relief along its trace, and curved fault segments visible in satellite imagery \citep{berberian1995,bahroudi2003,emami2010,tavani2018}
.

The Zagros region is seismically highly active, generating more than 50\% of Iran's earthquakes, representing nearly half of the Arabian Plate--Iran convergence, and is characterized by moderate to large earthquakes mainly associated with reverse and strike-slip faulting along major active structures such as the Main Zagros Fault (M.Z.F.), High Zagros Fault (H.Z.F.), Kazeron Fault (K.F.), and Mountain Front Fault (M.F.F.) (Fig.~\ref{fig:seismicity_map_zagros}b) \citep{mirzaei1998,tatar2002,hesami2003,engdahl2006,agard2011,fathian2021}
. The $M_w$ 7.3 Sar-e-Pol-e Zahab earthquake on 12 November 2017 was one of the most intense seismic events that posed a serious threat to people and structures in Iran's populated regions \citep{vetr2018}
. This event was one of the strongest to strike this region in the last hundred years, inducing an uplift of approximately 3500~km$^2$ between the H.Z.F. and the M.F.F., with the highest uplift of 70~cm south of the Miringe fault. In addition, subsidence occurred over an area of 1200~km$^2$ with a maximum of 35~cm at Vanisar village situated on the hanging wall of the H.Z.F. \citep{vajedian2018}
.

The distribution of seismic events with active faults and different structural units in the Zagros collision zone is shown in Fig.~\ref{fig:seismicity_map_zagros}b. The geographic distribution of seismic activity reveals high concentrations of tectonic activity within the Zagros Belt, extending from the northwest to the southeast. A significant number of seismic events occur with magnitudes between 1.5 and 5.0, indicating continuous deformation across the region. Of particular note are the stronger earthquakes with magnitudes between 5.0 and 7.3, which tend to occur along major tectonic boundaries. The Main Zagros Thrust (M.Z.T.) and the High Zagros Fault (H.Z.F.) are key structural elements in this context. The M.Z.T. marks the northeastern edge of the Zagros Belt and is considered the suture zone, exhibiting a notable concentration of seismic activity that suggests ongoing deformation and stress release. Similarly, the H.Z.F., located within the Zagros Belt, displays a clear pattern of seismicity, highlighting its role in accommodating the crustal shortening produced by the collision of the Arabian Plate with the Central Iran Block. The spatial distribution of earthquakes shows a high correlation with the dominant structural features of the area, particularly the Main Zagros Thrust and the High Zagros Fault.

\section{Brief Overview of Statistical Methods}\label{sec:methods}

\subsection{Gutenberg--Richter Relationship (\textit{b}-value) and Completeness Magnitude ($M_c$)}\label{subsec:bvalue}
The magnitude distribution of earthquakes is typically described by the Gutenberg--Richter law \citep{gutenberg1944}
which is expressed as:
\begin{equation}
    \log_{10} N(M) = a - bM
    \label{eq:gr}
\end{equation}
where $N$ is the number of earthquakes with a magnitude greater than or equal to $M$, $a$ is a positive constant that reflects the seismic activity level of the region, and $b$ (the $b$-value) indicates the balance between the number of large and small earthquakes in a given area. Maximum likelihood estimation (MLE) and least squares (LS) are two methods commonly used to estimate the $b$-value. In this study, the $b$-value was obtained using the maximum likelihood method \citep{aki1965}
. The equation used to determine the $b$-value is: 
\begin{equation}
    b = \frac{\log_{10} e}{\bar{M} - (M_c - \Delta M/2)}
    \label{eq:b_mle}
\end{equation}
where $\bar{M}$ is the average magnitude of the events in the analysis window, $M_c$ is the magnitude of completeness, and $\Delta M = 0.1$ is the magnitude binning interval. Selecting the appropriate method for $M_c$ estimation is a critical aspect of $b$-value analysis. We employed the widely accepted Maximum Curvature (MAXC) method \citep{wiemer2000}
to compute $M_c$. This procedure estimates $M_c$ by locating the peak of the first derivative of the frequency--magnitude distribution (FMD); in practical terms, $M_c$ is the most frequently occurring magnitude in the non-cumulative FMD. Because the $b$-value is time-dependent, its standard deviation $\delta b$ is calculated using the formulation of \citet{shi1982}
:
\begin{equation}
    \delta b = 2.30\, b^2 \sqrt{\frac{\displaystyle\sum_{i=1}^{n}(M_i - \bar{M})^2}{n(n-1)}}
    \label{eq:b_err}
\end{equation}
where $n$ denotes the total number of events and $M_i$ is the magnitude of the $i$-th earthquake.

\subsection{Fractal Correlation Dimension ($D_c$-value)}\label{subsec:dcvalue}
The fractal dimension ($D_c$-value) is a measure of spatial clustering that quantifies the distribution of earthquake hypocenters \citep{grassberger1983}
. The $D_c$-value for each cell is determined via the correlation integral method \citep{yadav2012}
. The Kriging method was employed to perform interpolation using the latitude and longitude coordinates of the center of each cell. The correlation integral is defined as:
\begin{equation}
    C(r) = \frac{2}{n(n-1)} \sum_{i=1}^{n-1}\sum_{j=i+1}^{n} H\!\left(r - |X_i - X_j|\right)
    \label{eq:corr_integral}
\end{equation}
where $r$ is the inter-epicenter distance, $X_i$ and $X_j$ are the spatial coordinates of the $i$-th and $j$-th epicenters, and $H$ is the Heaviside step function. $C(r)$ is the correlation function that counts the fraction of epicenter pairs separated by a distance less than $r$. If the spatial distribution of earthquake epicenters exhibits fractal characteristics, the following power-law relation holds:
\begin{equation}
    C(r) \propto r^{D_c}
    \label{eq:fractal}
\end{equation}
where $D_c$ is the fractal correlation dimension. The slope of the log--log plot of $C(r)$ versus $r$ yields $D_c$.

\subsection{Differential Stress Analysis}\label{subsec:stress}
The estimation of differential stress ($\Delta\sigma = \sigma_1 - \sigma_3$) from earthquake catalog data relies on the well-established empirical relationship between the Gutenberg--Richter $b$-value and the prevailing stress conditions. Numerous field observations and laboratory experiments demonstrate that the $b$-value systematically decreases as differential stress increases. This behavior reflects the suppression of small crack formation under elevated stress and the preferential growth of larger, critically oriented fractures \citep{amitrano2003}
. \citet{scholz2015}
applied a simplified frictional-resistance framework to evaluate lithospheric stress and examined spatial variations in the $b$-value across diverse tectonic settings and depths, refining earlier laboratory-based correlations. 
Similarly, field-based results indicate that the $b$-value declines with increasing applied stress in both continental and subduction environments, highlighting its sensitivity to depth and focal mechanism. Consequently, variations in the $b$-value can be interpreted as indicators of the underlying stress field. The methodological procedure is summarized as:
\begin{equation}
    \sigma_1 - \sigma_3 \; [\text{MPa}] = \frac{10^{(1-b)}}{c}
    \label{eq:diff_stress}
\end{equation}
where $(\sigma_1 - \sigma_3)$ is the stress difference between the maximum and minimum principal stresses in megapascals, and $c$ is an empirical constant \citep{scholz2015}
.

Spatial interpolation of the $b$-value, $D_c$-value, and differential stress was carried out using ordinary Kriging with a uniform grid resolution of $0.25^\circ$. A consistent grid configuration was applied to all parameters to maintain spatial comparability and ensure coherence among the resulting distribution maps. Fault traces from regional geological maps were included only as reference layers to preserve the structural framework and alignment with the main tectonic trends. To enable direct comparison between parameters, all interpolated datasets were resampled onto a common grid and standardized within a unified coordinate system. This approach allows consistent spatial overlay of the seismic indicators without introducing additional interpretative bias, providing a coherent dataset for subsequent spatial analysis.

\section{Seismic Dataset}\label{sec:dataset}
The data obtained from the Iranian Seismological Center (IRSC) were used to analyze the seismicity parameters. The dataset contains information on the date of occurrence, number of events, depth (hypocenter), and magnitude of the earthquakes. The earthquakes considered in this research have a magnitude of 1.5 and above, spanning a long observation period from January~1, 2006, to October~31, 2024. The dataset compiled and updated by the IRSC comprises a total of 40,731 unique seismic events. The seismic events considered in the present research are reported in Table~\ref{tab:dataset}.

\begin{table}[ht!]
\centering
\caption{Summary of the IRSC earthquake dataset used in this study.}
\label{tab:dataset}
\begin{tabular}{lccc}
\hline
\textbf{Year} & \textbf{Number of events} & \textbf{Magnitude range} & \textbf{Depth range (km)} \\
\hline
2006 & 913   & 1.5--6.1 & 0.1--36.0 \\
2007 & 1343  & 1.5--5.1 & 0.1--36.0 \\
2008 & 1549  & 1.5--6.0 & 0.3--36.0 \\
2009 & 1507  & 1.5--5.1 & 0.4--36.0 \\
2010 & 1969  & 1.5--6.1 & 0.1--36.0 \\
2011 & 1481  & 1.5--5.3 & 0.5--36.0 \\
2012 & 1795  & 1.5--5.5 & 1.2--45.0 \\
2013 & 2533  & 1.5--6.3 & 1.0--36.0 \\
2014 & 3098  & 1.5--6.2 & 0.7--36.0 \\
2015 & 1670  & 1.5--5.3 & 0.2--28.0 \\
2016 & 2246  & 1.5--5.4 & 0.1--27.4 \\
2017 & 3526  & 1.5--7.3 & 0.4--33.5 \\
2018 & 6434  & 1.5--6.4 & 1.0--28.0 \\
2019 & 2876  & 1.5--5.8 & 1.6--28.0 \\
2020 & 2082  & 1.5--5.7 & 1.0--28.0 \\
2021 & 2087  & 1.5--6.3 & 1.2--28.0 \\
2022 & 1403  & 1.5--6.1 & 0.4--28.0 \\
2023 & 1218  & 1.5--5.6 & 1.9--28.0 \\
2024 & 1001  & 1.5--4.9 & 0.7--28.0 \\
\hline
\end{tabular}
\end{table}

\begin{figure*}[ht!]
    \centering
    \includegraphics[width=\textwidth]{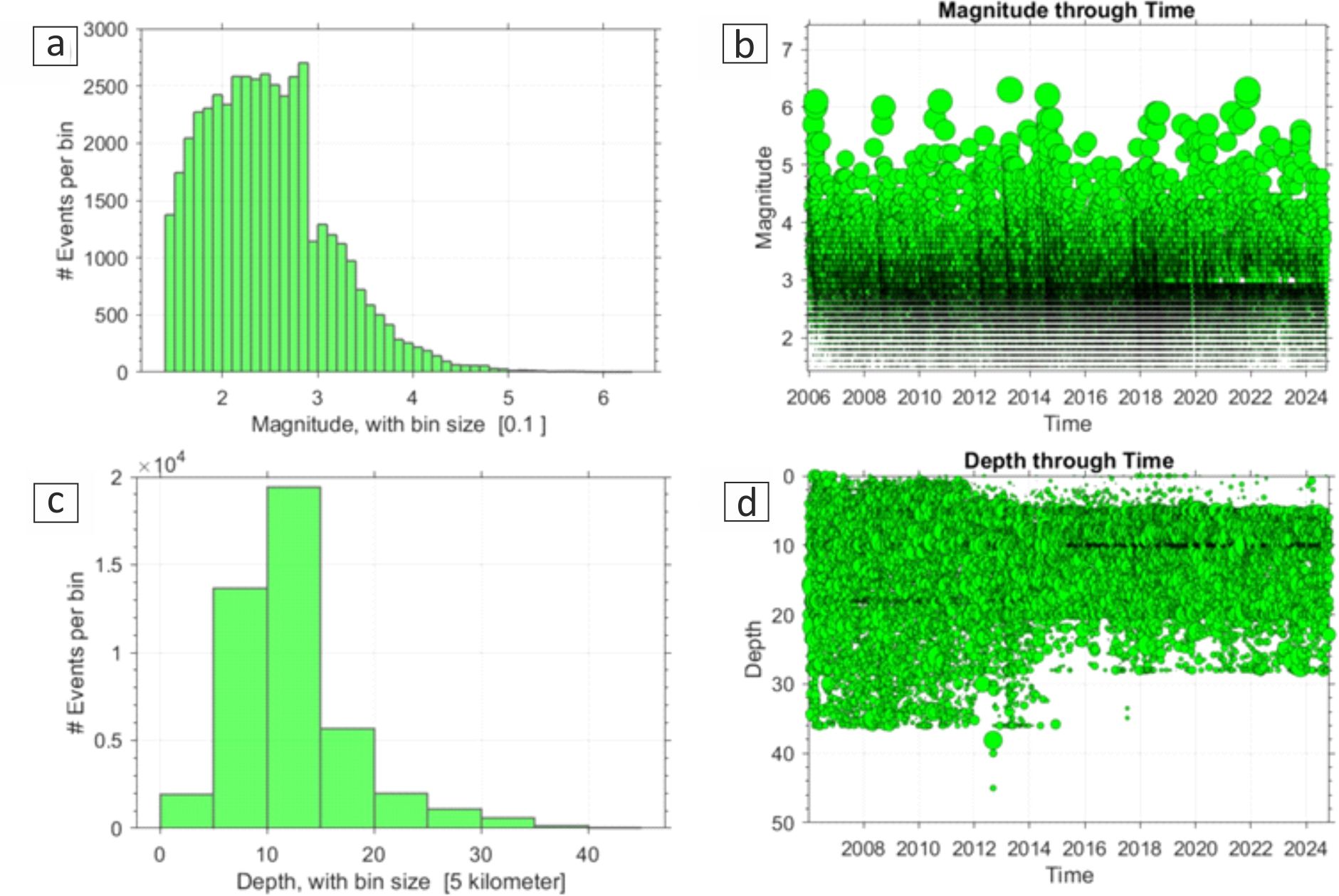}
    \caption{(a,~b) Magnitude distribution of events with respect to time. (c,~d) Depth distribution of events with respect to time.}
    \label{fig:catalog_stats}
\end{figure*}

Figure~\ref{fig:catalog_stats} provides insights into the depth and magnitude distribution of earthquakes and their temporal variations within the study area. The majority of earthquakes have magnitudes ranging between 2.0 and 3.0, followed by events with magnitudes between 3.0 and 4.0 (Fig.~\ref{fig:catalog_stats}a). Temporal analysis indicates that earthquakes with magnitudes between 2.0 and 5.0 have occurred regularly throughout the observation period, whereas events with magnitudes of 5.0 and greater have occurred intermittently since 2006 (Fig.~\ref{fig:catalog_stats}b). This pattern suggests persistent seismic activity in the region, with occasional moderate-to-strong earthquakes reflecting the ongoing accumulation and release of tectonic stress. The seismic activity in the region is confined mostly within the shallow to mid-upper crust, peaking at depths between 5 and 20~km with maximum activity recorded between 10 and 15~km (Fig.~\ref{fig:catalog_stats}c). The temporal distribution of earthquake depths indicates that seismic events have consistently occurred at depths between 5 and 20~km throughout the study period (Fig.~\ref{fig:catalog_stats}d). This is a standard depth profile for an active continental tectonic environment, such as fold-and-thrust belts or transform zones like the Zagros region, where brittle failure mechanisms dominate the upper crust. The sudden drop in seismic activity below approximately 25~km likely signals the depth of the brittle--ductile transition zone, where deformation tends to become more aseismic.

Homogenization of earthquake catalogues is a crucial step in seismic hazard assessment, as it ensures consistency in magnitude scales and enables reliable comparison across studies \citep{scordilis2006}
. In the original IRSC catalogue, earthquake magnitudes are reported in the local $M_N$ scale; therefore, to achieve a uniform dataset, all $M_N$ values were converted to moment magnitude ($M_w$) using established empirical relationships \citep{karimiparidari2013}
, allowing for a consistent and comparable seismic catalogue for further analysis. Moreover, the catalog utilized in this research has been declustered through the algorithm developed by \citep{gardner1974}
to enhance the dataset's reliability and facilitate stable estimations of seismic parameters. This method effectively eliminates aftershocks and foreshocks, enabling the distinction of independent mainshock events, which is essential for obtaining an unbiased estimation of seismic parameters.

\section{Results and Discussion}\label{sec:results}

\subsection{Frequency--Magnitude Distribution and Catalog Completeness}\label{subsec:fmd_completeness}
\begin{figure}[ht!]
    \centering
    \includegraphics[width=0.7\textwidth]{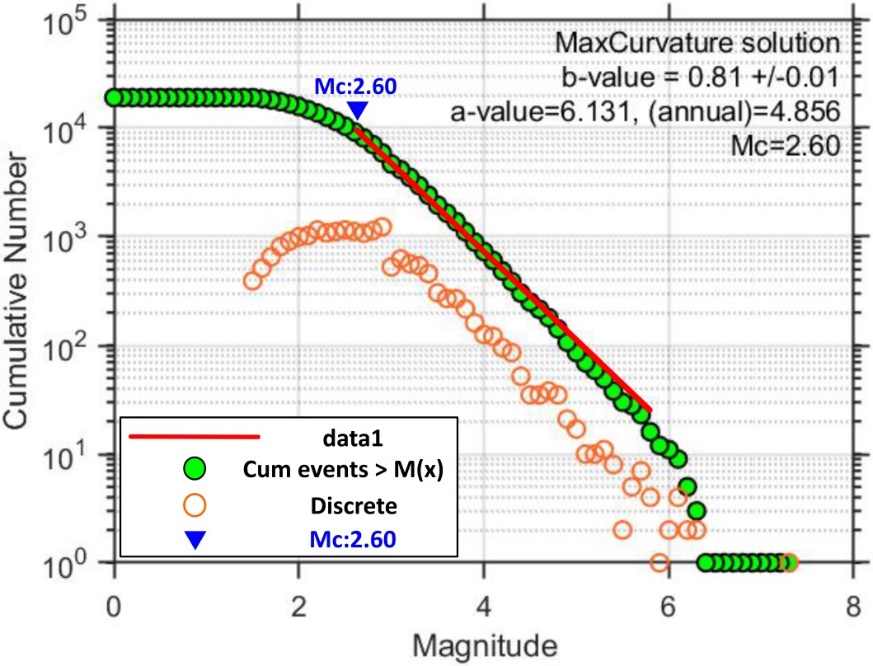}
    \caption{The cumulative frequency--magnitude distribution (FMD) in the study region based on the IRSC catalog from January~1, 2006, to October~31, 2024. The red line represents the Gutenberg--Richter equation fit.}
    \label{fig:fmd}
\end{figure}

The baseline evaluation of the frequency--magnitude distribution (FMD) represents the cornerstone of any statistical seismology investigation, ensuring that subsequent spatial and temporal parameter variations are not biased by catalog incompleteness or network-related detection limits. The earthquake catalog compiled from the Iranian Seismological Center (IRSC) database, covering the temporal window from January 1, 2006, to October 31, 2024, was subjected to a rigorous magnitude of completeness ($M_c$) assessment using the Maximum Curvature (MAXC) method of \citet{wiemer2000}. By identifying the peak of the non-cumulative FMD, the MAXC approach established a stable completeness threshold at $M_c = 2.60$ for the entire study area (Fig.~\ref{fig:fmd}). This threshold represents a high detection capability for a regional seismic network covering a geographically expansive and complex mountain belt. 

Using the maximum likelihood estimation (MLE) method proposed by \citet{aki1965}, the overall $b$-value for the compiled catalog was calculated as $0.81 \pm 0.01$ (Fig.~\ref{fig:fmd}), with a corresponding seismic productivity value (or $a$-value) of 6.32 (as estimated via the Gutenberg--Richter relation). A regional $b$-value of 0.81 is significantly lower than the global tectonic average of approximately 1.0 \citep{frohlich1993, kagan1999}, reflecting a state of elevated differential stress and a higher proportion of moderate-to-large events relative to small earthquakes across the Zagros collision zone. This low $b$-value matches findings from other highly active, compressional settings in Iran, such as the Alborz mountain belt \citep{tourani2024}, and is physically consistent with rapid tectonic loading under the Arabia--Eurasia convergence. 

To contextualize these findings within the historical stress evolution of the Zagros Orogenic Belt, we compare our results with decadal trends reported in the literature. Mousavi-Bafrouei et al.\ (2014) analyzed temporal fluctuations of the average $b$-value across Iran and documented a progressive decline within the Zagros seismogenic zone, reporting values of 1.42 for the 1900--1963 interval, 1.29 for 1964--1996, and 0.96 for 1997--2012. Subsequent retrospective studies by Mousavi (2016) and \citet{mousaviyan2022} confirmed the physical reality of this decadal trend. Our calculated $b$-value of $0.81 \pm 0.01$ using recent catalog data (2006--2024) indicates that this decline has continued over the last two decades. Physically, this progressive $b$-value reduction represents an accelerating tectonic loading phase, in which the seismogenic crust has transitioned from distributed, low-stress fracturing to a mature state of high differential stress concentration and widespread fault locking.

\subsection{Spatial Mapping of the $b$-value: Delineation of Locked Fault Segments and Asperities}\label{subsec:spatial_b}
\begin{figure*}[ht!]
    \centering
    \includegraphics[width=\textwidth]{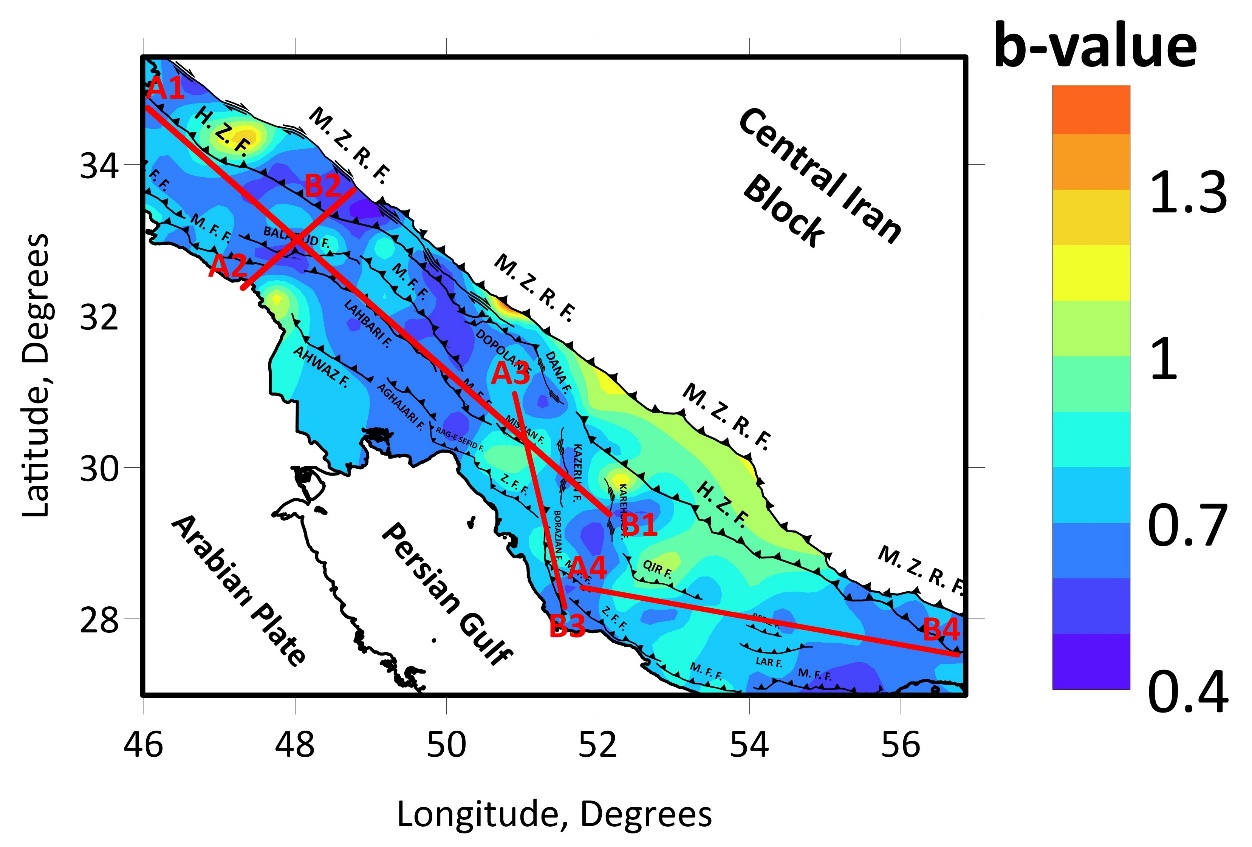}
    \caption{Spatial variation of $b$-value in the Zagros Mountains according to the IRSC catalogue. The red solid lines indicate the locations of the cross-sections shown in Figure~\ref{fig:bvalue_sections}.}
    \label{fig:bvalue_map}
\end{figure*}

While the catalog-wide $b$-value of 0.81 provides a macro-scale tectonic characterization, resolving the spatial heterogeneity of the $b$-value is critical for mapping localized stress concentrations and delineating potential seismic asperities. Figure~\ref{fig:bvalue_map} displays the spatial distribution of the $b$-value across the Zagros region, interpolated using ordinary Kriging on a uniform grid resolution of $0.25^\circ \times 0.25^\circ$. The spatial field exhibits pronounced variations, with $b$-values ranging from approximately 0.4 to 1.3 across different structural domains. The most striking feature of the map is the systematic, first-order alignment of low $b$-value anomalies ($0.4$--$0.7$) along the trace of major fault systems and regional structural boundaries.

According to geomechanical models and laboratory experiments, the $b$-value scales inversely with applied differential shear stress ($\sigma_1 - \sigma_3$), meaning that low $b$-values trace areas of critical tectonic loading and high-stress accumulation \citep{scholz1968, schorlemmer2005, scholz2015}. In the Zagros, these low $b$-value zones are concentrated along key reverse and thrust fault networks, notably the Mountain Front Fault (M.F.F.), the central and southern segments of the High Zagros Fault (H.Z.F.), and the Main Zagros Reverse Fault (M.Z.R.F.) (Fig.~\ref{fig:bvalue_map}). Additionally, localized low $b$-value patches are observed near the Balarud Fault and the southern terminations of the Ahwaz, Lahbari, Rag Sefid, Aghajari, Kazerun, Lar, and Beriz faults. These structures accommodate the oblique northward motion of the Arabian Plate relative to the Eurasian Plate. According to \citet{demets2010}, this plate convergence occurs at a rate of approximately 31~mm/yr, while GPS-based studies propose slightly lower convergence rates of around 21~mm/yr \citep{mcclusky2000, vernant2004}. This plate shortening is partitioned into fault-parallel thrusting and strike-slip faulting. The concentration of low $b$-values along these major tectonic boundaries suggests that these fault segments are strongly locked and acting as major asperities that accumulate elastic strain under high differential stress, presenting a high potential for generating future moderate-to-large earthquakes.

\subsection{Depth-Dependent $b$-value Stratification: Brittle-Ductile Transitions and Stress Localization}\label{subsec:depth_b}
\begin{figure*}[ht!]
    \centering
    \includegraphics[width=\textwidth]{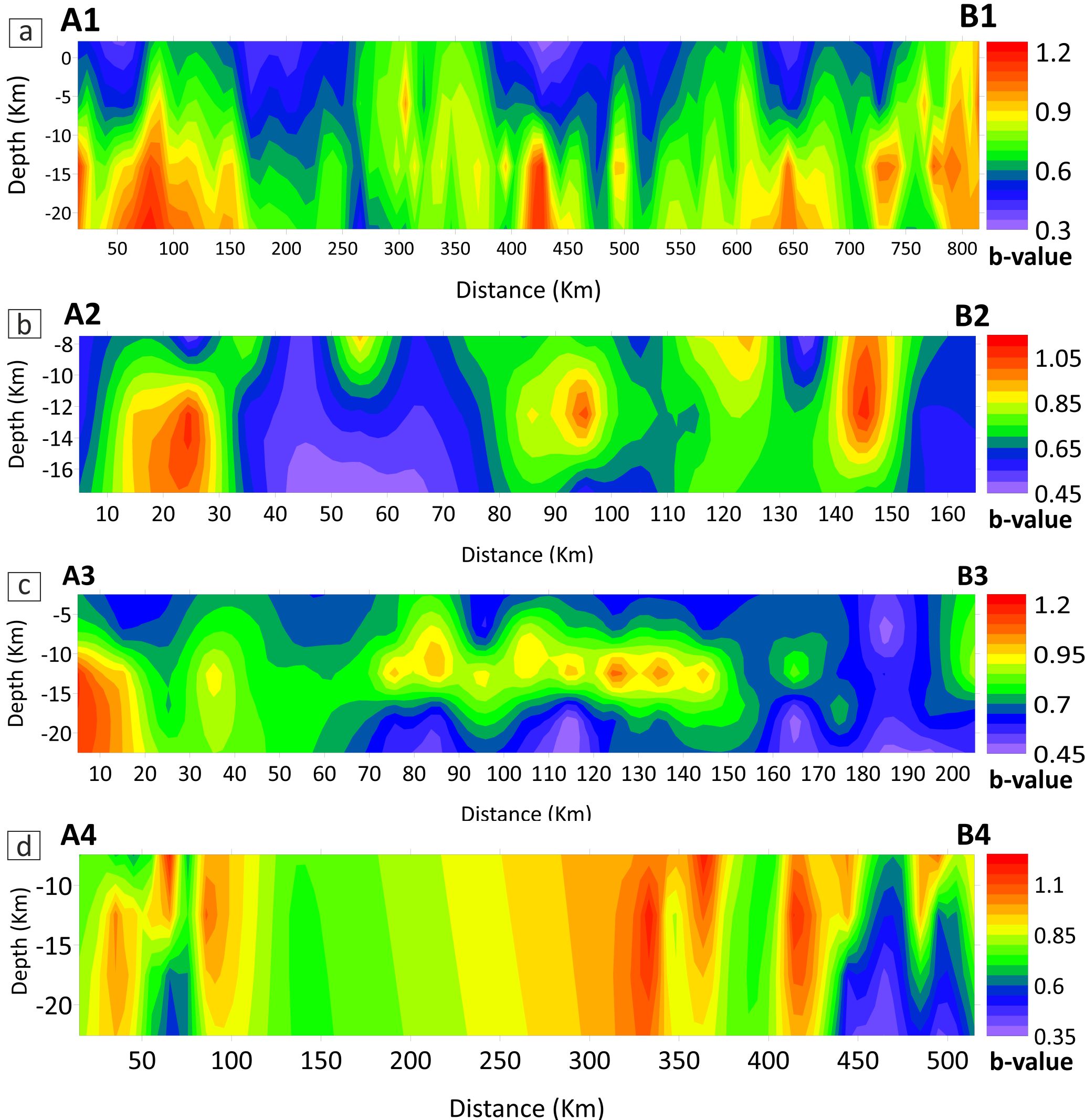}
    \caption{Cross-sectional profiles (A1--B1 to A4--B4) showing the $b$-value distribution at the locations indicated in Figure~\ref{fig:bvalue_map}.}
    \label{fig:bvalue_sections}
\end{figure*}

To characterize the vertical stress field and understand how tectonic loading varies within the seismogenic column, we constructed four depth-dependent cross-sectional profiles (A1--B1 to A4--B4) across key active domains of the Zagros Orogenic Belt (Fig.~\ref{fig:bvalue_sections}). These profiles reveal a pronounced depth-dependent stratification of the $b$-value, pointing to a strong vertical control on stress accumulation.

Along the northwesternmost profile (A1--B1), very low $b$-values ($b < 0.6$) are concentrated within the shallowest $\sim$10~km of the crust (Fig.~\ref{fig:bvalue_sections}a). This concentration indicates a shallow zone of high differential stress and localized strain accumulation, representing a high seismic hazard potential within the brittle upper crust. Moving to the central-northern sector, the A2--B2 profile exhibits a more heterogeneous and fragmented vertical structure. In this section, low $b$-value patches are confined to discrete depths of $\sim$8--16~km (Fig.~\ref{fig:bvalue_sections}b), suggesting that tectonic stress is distributed across multiple sub-parallel structures or offset fault segments rather than concentrated on a single shallow interface. 

The A3--B3 profile, crossing the central Zagros, represents one of the most critical structural segments in our dataset. It displays a vertically continuous, high-stress anomaly characterized by $b$-values of $\sim$0.45--0.7 extending from shallow depths down to approximately 15--20~km (Fig.~\ref{fig:bvalue_sections}c). This vertical continuity indicates strong mechanical coupling and uniform stress accumulation across the entire brittle seismogenic layer, suggesting a mature fault patch capable of generating large-scale crustal ruptures. In contrast, the southeasternmost profile (A4--B4) exhibits generally higher and more variable $b$-values (Fig.~\ref{fig:bvalue_sections}d). This pattern reflects a transition toward more distributed, ductile deformation and structural complexity as the belt approaches its southeastern termination (near the Minab Fault zone and the Makran subduction transition). 

Across all four profiles, a consistent geomechanical pattern emerges: low $b$-values ($b < 0.6$) are largely restricted to the upper $\sim$10--15~km of the crust, corresponding to the seismogenic brittle layer. Below this depth, the $b$-value increases, reflecting a transition to ductile deformation, elevated pore-fluid pressures, and a reduction in differential shear stress. This vertical stratification confirms that the shallow brittle crust along the main fault zones is the primary reservoir of strain accumulation and seismic energy release in the Zagros Orogenic Belt.

\subsection{Spatial and Depth-Dependent Fractal Dimension ($D_c$): Evaluating Fault Heterogeneity and Strain Localization}\label{subsec:dc_analysis}
\begin{figure*}[ht!]
    \centering
    \includegraphics[width=\textwidth]{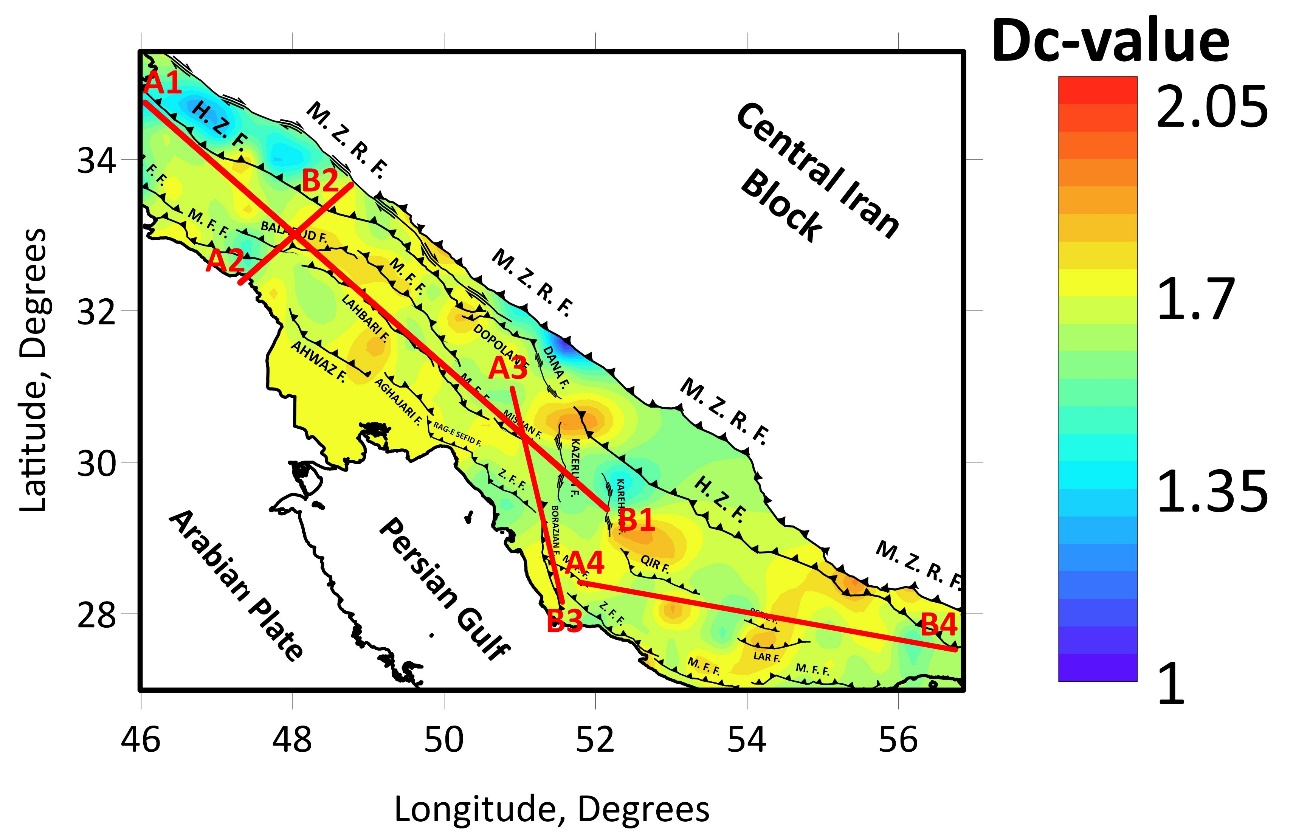}
    \caption{Spatial variation of $D_c$-value in the Zagros Mountains according to the IRSC catalogue. The red solid lines indicate the locations of the cross-sections shown in Figure~\ref{fig:dc_sections}.}
    \label{fig:dc_map}
\end{figure*}

The correlation fractal dimension ($D_c$-value) provides a geometric perspective on the seismotectonic state, quantifying the spatial clustering properties and structural complexity of the epicentral distribution. While $D_c \approx 1.0$ indicates seismicity closely aligned along linear fault segments, $D_c \approx 2.0$ represents a planar distribution or space-filling seismicity, and values between these limits describe varying degrees of fracturing and fault segmentation \citep{grassberger1983, yadav2012a, ozturk2019}. In the Zagros region, the calculated spatial $D_c$-values range from 1.0 to 2.05 (Fig.~\ref{fig:dc_map}), indicating a moderate-to-high spatial heterogeneity.

The spatial distribution map (Fig.~\ref{fig:dc_map}) reveals that high $D_c$-values ($\ge 1.5$) dominate a large portion of the Zagros Orogenic Belt. This high spatial complexity reflects the highly fractured, segmented, and structurally heterogeneous nature of the fold-and-thrust belt, which comprises multiple sub-parallel fault strands and complex folding structures. Notably, zones of high $D_c$-values show a strong spatial correspondence with the low $b$-value zones described in Section~\ref{subsec:spatial_b}. This co-location is physically meaningful: in highly active, compressional settings, stress concentration occurs along complex fault networks with high material heterogeneity, leading to the simultaneous occurrence of low $b$-values (elevated stress) and high $D_c$-values (complex, space-filling fracturing) \citep{bai2021, tourani2024}.

\begin{figure*}[ht!]
    \centering
    \includegraphics[width=\textwidth]{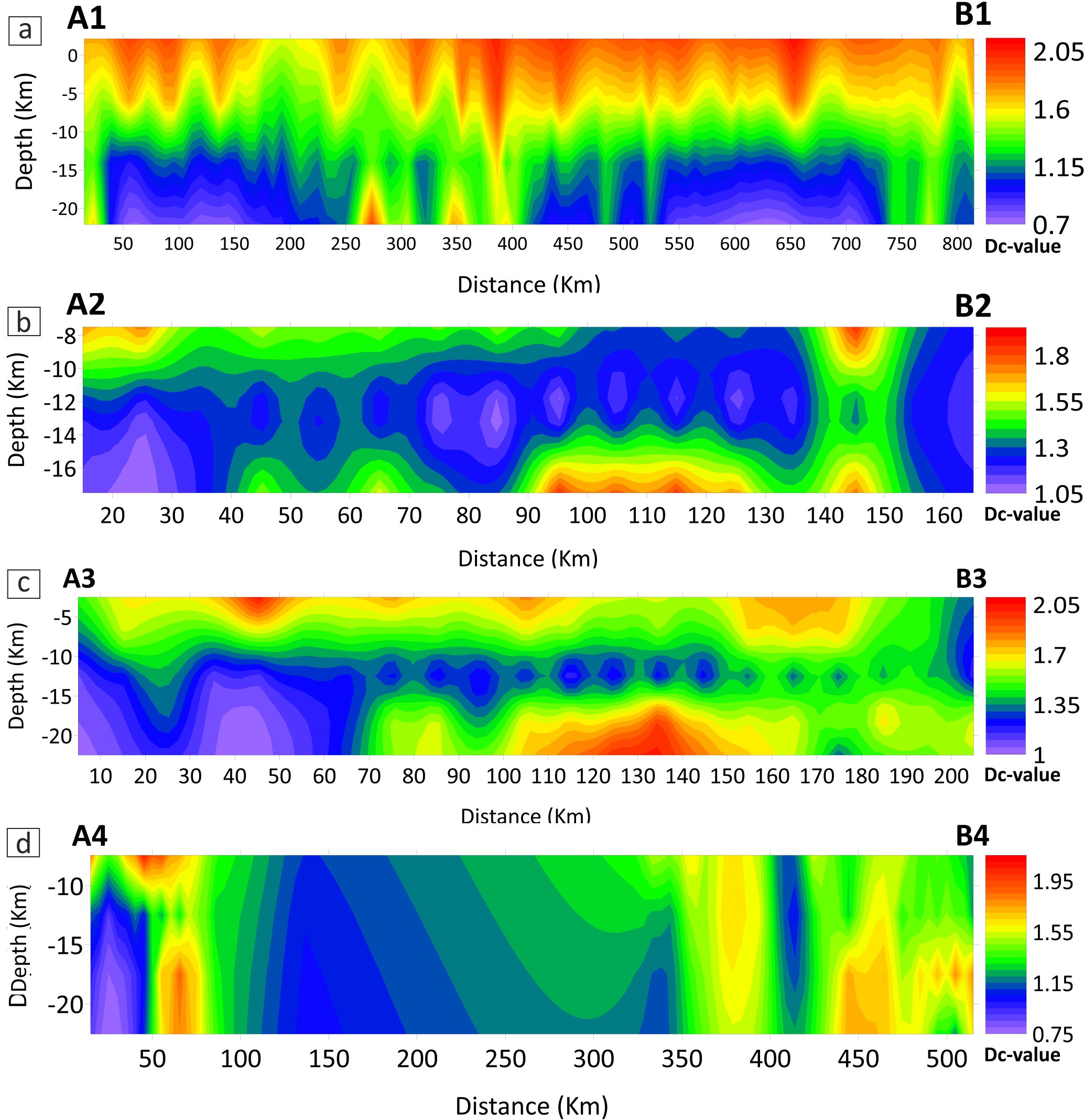}
    \caption{Cross-sectional profiles (A1--B1 to A4--B4) showing the $D_c$-value distribution at the locations indicated in Figure~\ref{fig:dc_map}.}
    \label{fig:dc_sections}
\end{figure*}

To investigate how this spatial complexity behaves at depth, we analyzed the $D_c$ depth-dependent cross sections (Fig.~\ref{fig:dc_sections}). Along profile A1--B1, high $D_c$ values of 1.6--2.05 are concentrated in the upper $\sim$5~km of the crust, while values below 1.15 dominate the 10--20~km depth range (Fig.~\ref{fig:dc_sections}a). This vertical variation represents a transition from shallow, highly fragmented deformation (associated with complex folding and secondary faulting in the sediment cover) to more localized, planar slip along mature fault zones at depth. Profile A2--B2 exhibits lower, more uniform $D_c$-values ($\le 1.3$) throughout the crustal column (Fig.~\ref{fig:dc_sections}b), indicating a simpler, more organized fault geometry and lower spatial complexity.

In the central A3--B3 profile, low $D_c$-values ($D_c < 1.35$) spatially coincide with low-to-intermediate $b$-values ($0.7$--$1.0$) at depths of 10--20~km (Fig.~\ref{fig:dc_sections}c). This co-location occurs along the deep segments of the Mishan, Borazjan, Mountain Front, and Zagros Foredeep faults, highlighting these segments as highly localized, critically stressed asperities that are accumulating strain on discrete fault surfaces. Similarly, the A4--B4 profile shows that zones of low $D_c$-values ($D_c < 1.2$) correlate with $b$-values below 1.0 (Fig.~\ref{fig:dc_sections}d), reflecting seismicity clustered along discrete, high-stress fault strands. In summary, the $D_c$ profiles document that while the upper sediment cover and shallow crust exhibit high spatial complexity (high $D_c$), seismicity at seismogenic depths along major faults tends to localize onto simpler, discrete planes (lower $D_c$), indicating a transition to coherent fault rupture behavior.

\subsection{Crustal Differential Stress Distribution and Its Scaling with Seismicity Parameters}\label{subsec:stress_analysis}
\begin{figure*}[ht!]
    \centering
    \includegraphics[width=\textwidth]{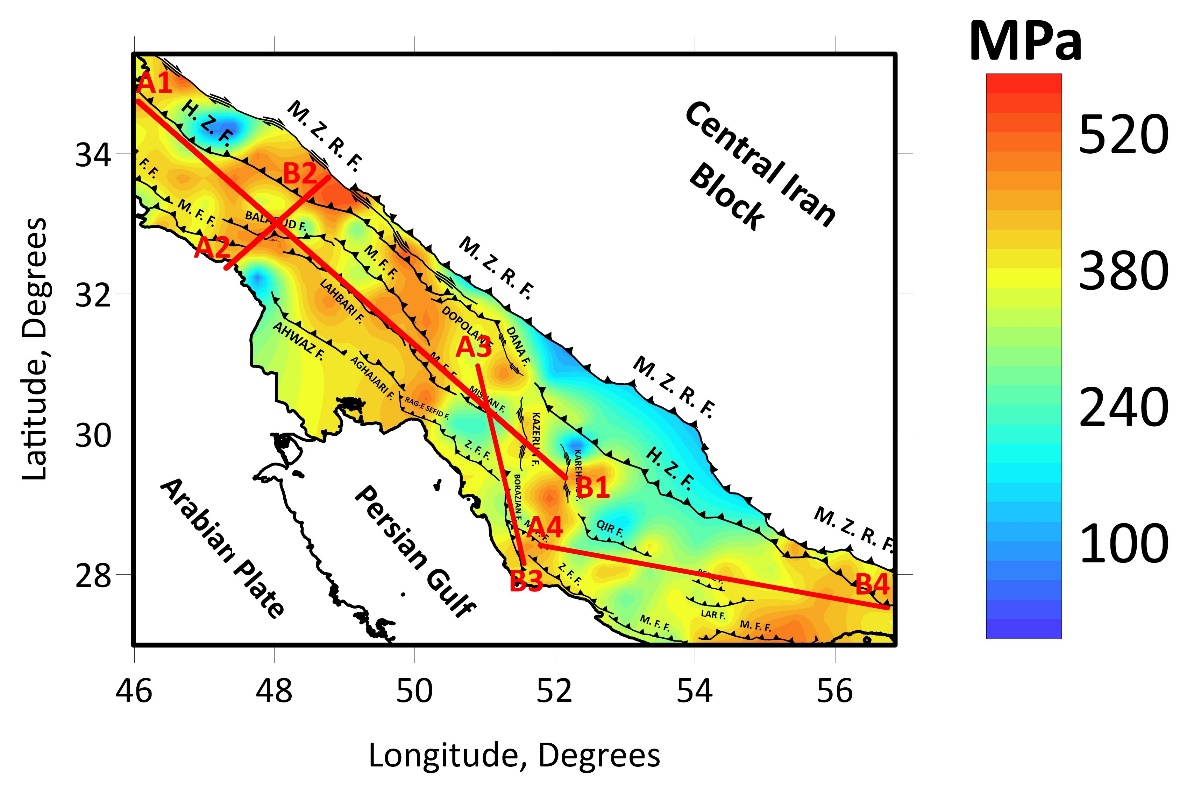}
    \caption{Spatial variation of differential stress ($\sigma_1 - \sigma_3$) in the Zagros Mountains according to the IRSC catalogue. The red solid lines indicate the locations of the cross-sections shown in Figure~\ref{fig:stress_sections}.}
    \label{fig:stress_map}
\end{figure*}

To quantify the stress state of the crust, we estimated the regional differential stress ($\sigma_1 - \sigma_3$) using the empirical formulation proposed by \citet{scholz2015}. This model establishes a quantitative link between the Gutenberg--Richter $b$-value and the prevailing differential shear stress, with low $b$-values acting as a proxy for high differential stress. As shown in Figure~\ref{fig:stress_map}, the spatial distribution of differential stress across the Zagros Mountains varies between approximately 100 and 520~MPa, indicating a highly heterogeneous stress field.

Across a large portion of the belt, the differential stress is concentrated at or above 520~MPa (Fig.~\ref{fig:stress_map}), reflecting the high tectonic loading associated with the ongoing Arabia--Eurasia collision. This region-wide stress field exhibits a clear inverse relationship with the spatial $b$-value map (Fig.~\ref{fig:bvalue_map}), validating the Scholz (2015) stress scaling relationship. The highest differential stresses ($\ge 520$~MPa) systematically overlap with the lowest $b$-value zones ($0.4$--$0.7$) along the Mountain Front Fault, High Zagros Fault, and Main Zagros Reverse Fault, confirming that these segments are under extreme tectonic loading and are close to failure.

\begin{figure*}[ht!]
    \centering
    \includegraphics[width=\textwidth]{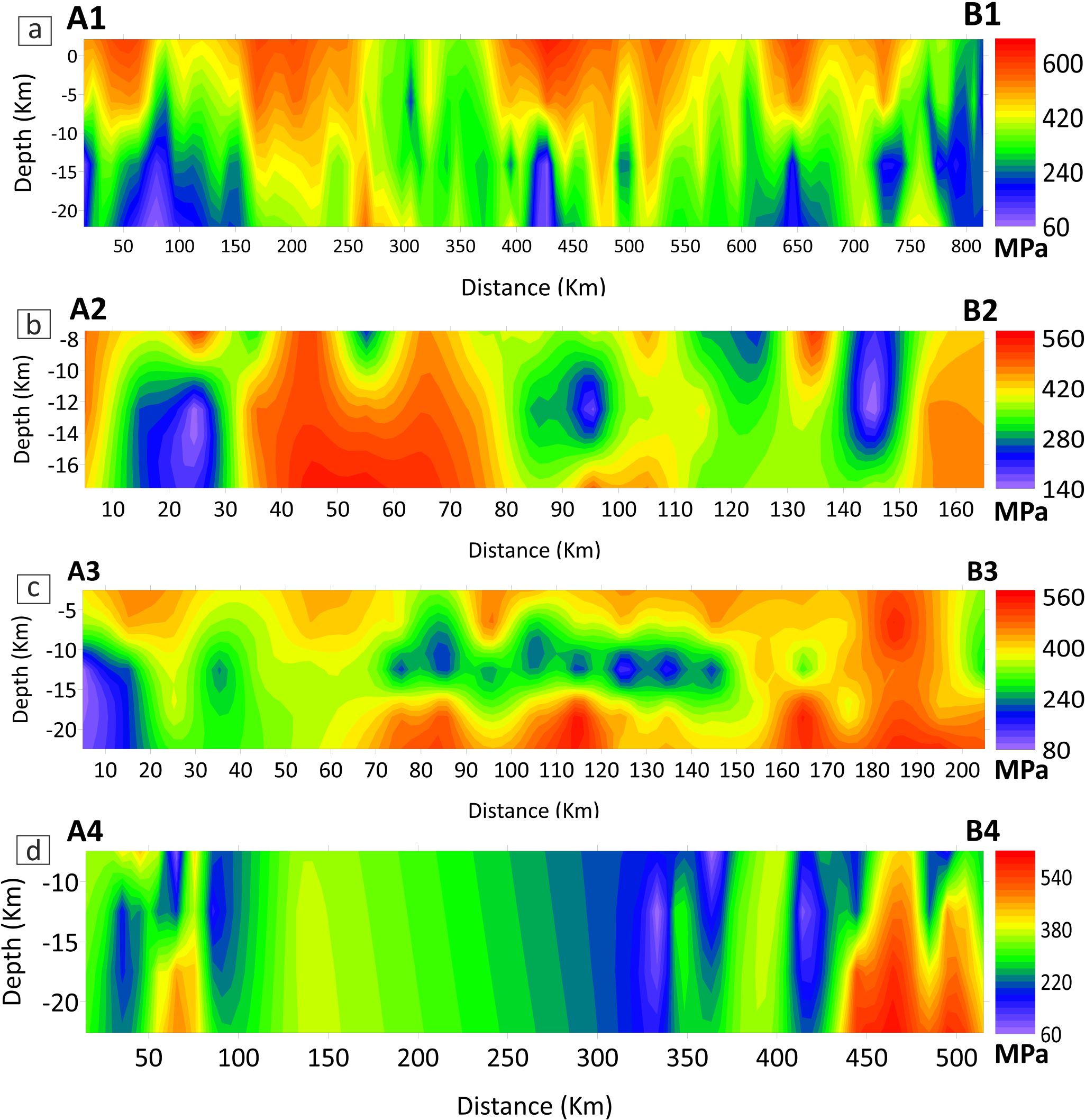}
    \caption{Cross-sectional profiles (A1--B1 to A4--B4) showing the differential stress ($\sigma_1 - \sigma_3$) distribution at the locations indicated in Figure~\ref{fig:stress_map}.}
    \label{fig:stress_sections}
\end{figure*}

This geomechanical coupling is further supported by the differential stress cross-sectional profiles (A1--B1 to A4--B4; Fig.~\ref{fig:stress_sections}). These profiles present an almost exact mirror image of the $b$-value cross sections (Fig.~\ref{fig:bvalue_sections}). High differential stress zones (exceeding 500~MPa) consistently correspond to low $b$-value patches across all four profiles. Along the A1--B1 and A3--B3 profiles, these high-stress zones are concentrated in the upper 10--15~km of the crust, whereas the A2--B2 profile shows a more fragmented stress field at 8--16~km depth, and the A4--B4 profile displays lower stress levels toward the southeast. This consistent spatial and depth correspondence between two independent parameters confirms the geomechanical validity of our results, confirming that the Gutenberg--Richter $b$-value serves as a reliable proxy for tracking the spatial and vertical evolution of the crustal stress field in the Zagros Mountains.

\subsection{Synthesis: Geomechanical Coupling and Seismic Hazard Assessment in the Zagros Orogenic Belt}\label{subsec:synthesis_zagros}
The multi-parameter analysis presented in this study reveals a coherent and geomechanically consistent description of the crustal stress state and seismic hazard along the Zagros Orogenic Belt. The spatial and depth convergence of the $b$-value, fractal dimension ($D_c$), and differential stress ($\sigma_1 - \sigma_3$) is not a random statistical correlation, but rather represents the physical expression of localized strain accumulation and fault locking.

By integrating these parameters, we can construct a unified geomechanical model of the Zagros seismogenic crust. In the spatial domain, the co-location of low $b$-values ($0.4$--$0.7$), high $D_c$-values ($\ge 1.5$), and elevated differential stress ($\ge 520$~MPa) along the Mountain Front Fault, High Zagros Fault, and Main Zagros Reverse Fault identifies these structures as the primary loci of strain accumulation. The high $D_c$-values suggest that these fault zones are structurally complex and heterogeneous, while the low $b$-values and high differential stress indicate that this complexity is under critical tectonic loading. At depth, the cross-sectional profiles reveal that this critical stress state is concentrated within the upper $\sim$10--15~km of the crust, corresponding to the brittle seismogenic zone, while the localized decrease in $D_c$ at these depths suggests that seismicity is focusing onto discrete fault surfaces in preparation for future ruptures.

These findings have direct implications for seismic hazard assessment in western and southern Iran. The identified locked segments and high-stress asperities along the MFF, HZF, and MZRF represent the most likely sites for future moderate-to-large earthquakes, similar to the 2017 $M_w$ 7.3 Sarpol-e Zahab event. The progressive long-term decline of the regional $b$-value to 0.81 suggests that the Zagros crust is entering a stage of heightened criticality, with a growing likelihood of larger-magnitude events. Continuous monitoring of these spatio-temporal variations in $b$-value, $D_c$, and differential stress is therefore essential for updating intermediate-term forecasting models, refining regional hazard maps, and implementing effective earthquake risk reduction strategies in populated areas adjacent to these active fault systems.

The spatial and vertical integration of seismicity scaling parameters, and seismic velocity \citep{talebi2020} structures provides a multi-parameter view of the Zagros Orogenic Belt's mechanical state. Our findings reveal that the seismic behavior of the crust is not merely a product of fault kinematics, but is fundamentally governed by the vertical rheological architecture, specifically the presence of a mid-to-lower crustal felsic channel. The vertical seismic tomographic results \citep{talebi2020} identify a prominent low-velocity zone ($V_p$ anomalies at 10--25~km depth) that were interpreted as a felsic crustal layer. This LVZ serves as the primary ``governor'' of the crustal stress state. Based on our results, in regions where this felsic channel is well-developed, we observe a consistent geophysical signature:
\begin{enumerate}
    \item \textbf{Stress Dissipation (Low $\Delta\sigma$):} The differential stress values drop significantly (100--240~MPa) in these regions. The `soft' rheology of the felsic material, likely governed by high quartz content \citep{burgmann2008} and further exacerbated by hydrolytic weakening or high pore-fluid pressures \citep{griggs1965, koch1989, miranda2023}, prevents the accumulation of high elastic strain, resulting in the observed stress-limited seismic regime.
    \item \textbf{Fracture Scaling (High $b$-value):} Coincident high $b$-values indicate that the energy release is dominated by small-magnitude events. The crust in these zones yields at lower stress thresholds, preventing the formation of the large-scale asperities required for great earthquakes.
    \item \textbf{Spatial Complexity (High $D_c$-value):} The high $D_c$-values ($>1.7$) suggest a high fractal dimension of seismicity. This implies that deformation is accommodated through a spatially diffuse, three-dimensional network of fractures rather than being localized on a singular, rigid fault plane.
\end{enumerate}

The integration of these datasets also allows us to define two distinct seismotectonic regimes within the Zagros:
\begin{itemize}
    \item \textbf{The Northern/Central ``Locked'' Regime:} Characterized by higher velocities, low $b$-values, and exceptionally high differential stress. In these segments, the absence or thinning of the felsic channel leads to a ``stiffer'' crustal column. Tectonic convergence is stored as significant elastic strain within rigid, basement-involved structures. This creates a high-hazard environment where low $D_c$-values indicate high spatial localization, a precursor to major asperity-type seismic ruptures.
    \item \textbf{The Southern ``Decoupled'' Regime:} Characterized by the prominent felsic LVZ, high $b$-values, and low differential stress. Here, the felsic channel facilitates mechanical decoupling between the Arabian basement and the sedimentary cover. This decoupling results in a ``stress-limited'' state where deformation is distributed and dissipated through frequent, low-magnitude seismicity.
\end{itemize}

\section{Conclusions}\label{sec:conclusions}
In this study, the earthquake catalog of the Iranian Seismological Center (IRSC), covering the period 2006--2024, was used to construct a homogeneous seismic dataset for the Zagros Orogenic Belt. Based on this dataset, the spatial distribution of seismicity was analyzed and key seismotectonic parameters---including $b$-value, differential stress ($\sigma_1 - \sigma_3$), and fractal dimension ($D_c$-value)---were computed to investigate stress variations, clustering behavior, seismotectonic characteristics, and earthquake hazard potential across the region. The following main conclusions are drawn:

\begin{enumerate}

\item The $b$-value obtained in this study ($0.81 \pm 0.01$) represents a dominance of higher-magnitude events relative to smaller earthquakes, reflecting a high-stress state of the crust. This result indicates that the Zagros Orogenic Belt is tectonically active and currently under significant tectonic loading, suggesting elevated seismic hazard conditions in the region.

\item The spatial distribution of $b$-values demonstrates a strong first-order tectonic control, where persistently low $b$-value zones (0.4--0.7) are systematically aligned with major fault zones and structural boundaries. The consistent occurrence of low $b$-value anomalies along the Mountain Front Fault (M.F.F.), High Zagros Fault (H.Z.F.), and Main Zagros Reverse Fault (M.Z.R.F.) highlights these zones as critically stressed segments with enhanced tectonic loading and elevated seismic hazard.

\item The cross-sectional analysis reveals that $b$-value variations in the Zagros are strongly controlled by fault geometry, segmentation, and depth-dependent stress accumulation. Persistently low $b$-values ($<$0.6) are concentrated within the upper $\sim$10~km of the crust, indicating that the shallow brittle crust along major fault zones represents the primary zone of stress concentration and seismic energy release. Below this depth, increasing and more variable $b$-values suggest a transition toward more ductile deformation and reduced seismic sensitivity.

\item $D_c$-values ranging from 1.0 to 2.05, with a dominant distribution of relatively high values ($\sim$1.5 and above), together with their spatial coincidence with low $b$-value zones, indicate that the Zagros Mountains are characterized by highly heterogeneous and structurally complex fault zones. The cross-sectional profiles further show that high $D_c$-values dominate the upper crust, indicating intense deformation partitioning and complex fault-controlled seismicity within the brittle seismogenic layer, while localized decreases in $D_c$ with depth reflect more coherent rupture patterns in specific segments.

\item The differential stress ($\sigma_1 - \sigma_3$) distribution (100--520~MPa, with dominant values $\geq$520~MPa across much of the Zagros) indicates a region-wide state of high tectonic loading driven by the ongoing Arabia--Eurasia collision. The clear inverse relationship between differential stress and $b$-value confirms that low $b$-value zones correspond to high-stress regions. The cross-sectional analysis of differential stress shows a consistent inverse correspondence with $b$-value profiles, with both parameters systematically identifying the same critically stressed, fault-controlled regions. The additional agreement with high $D_c$-values further confirms that these areas are characterized by complex seismic clustering, elevated stress accumulation, and enhanced seismic hazard potential.

\item Collectively, the integrated use of $b$-value, differential stress ($\sigma_1 - \sigma_3$), and $D_c$-value is highly effective in delineating the most seismically hazardous segments of the Zagros Mountains. The hazardous areas identified correspond to critically stressed fault segments capable of generating destructive earthquakes, emphasizing the need for continuous monitoring, detailed seismic hazard assessment, and the implementation of effective mitigation and preparedness strategies across the region.

\end{enumerate}

\begingroup
\hyphenpenalty=10000
\exhyphenpenalty=10000
\bibliographystyle{elsarticle-harv}
\bibliography{mybibfile}
\endgroup

\end{document}